\documentclass[prb,a4paper,twocolumn,floatfix,showpacs,showkeys,amsmath,amssymb,nobibnotes,altaffilletter]{revtex4}

\usepackage{graphicx}
\usepackage{pslatex}
\usepackage{xspace}

\newcommand{\CIGSSe}{Cu(In,Ga)(S,Se)$_2$\xspace}
\newcommand{\CIGSe}{Cu(In,Ga)Se$_2$\xspace}

\newcommand{\degree}{$^{\circ}$}

\begin{document}

\preprint{}

\title{Band Bending Independent of Surface Passivation in ZnO/CdS/\CIGSSe
Heterojunctions and Cr/\CIGSSe Schottky Contacts}

\author{C.~Deibel}
\author{V.~Dyakonov}
\author{J.~Parisi}\email{Parisi@ehf.uni-oldenburg.de}
\affiliation{Department of Energy and Semiconductor Research,
Faculty of Physics, University of Oldenburg, 26111 Oldenburg, Germany}

\date{\today}

\begin{abstract}

We have employed admittance spectroscopy and deep-level transient spectroscopy
in order to investigate the electronic properties of ZnO/CdS/\CIGSSe
heterojunctions and Cr/\CIGSSe Schottky contacts. Our work concentrates on the
origin of an energy-distributed defect state commonly found in these systems.
The activation energy of the defect state addressed continuously shifts upon
air annealing or damp-heat treatment and is a valuable measure of the degree of
band bending of in \CIGSSe-based junctions. We demonstrate that the band
bending within the \CIGSSe layer, reported in the literature to become minimal
after air exposure, returns after the formation of either a Schottky contact or
a heterojunction. The above phenomenon turns out to be independent of a surface
passivation due to the CdS bath deposition.

\end{abstract}



\maketitle

Thin-film solar cells based on polycrystalline \CIGSSe chalcopyrite absorbers
yield a relatively high energy conversion efficienciy of 18.8\% on laboratory
scale devices~\cite{contreras99} and up to 12.5\% on large area
modules.~\cite{powalla02,probst01a} Still, some fundamental characteristics of
the above material system are not understood completely, one of them being the
diverse processes influencing the band bending of the heterojunction. A model
proposed by Rau et al.~\cite{rau99b} explains the comparatively strong band
bending of heterojunction devices by an absorber surface passivation due to the
subsequent CdS bath deposition. In this letter, we present admittance
spectroscopy and deep-level transient spectroscopy (DLTS) measurements on
ZnO/CdS/\CIGSSe heterojunctions and Cr/\CIGSSe Schottky contacts, showing that
the strong band bending derives from the process of contact formation at the
\CIGSSe-based junction independent of a CdS induced passivation.

The samples investigated were non-encapsulated ZnO/CdS/\CIGSSe solar cells and
Cr/\CIGSSe Schottky diodes. The fabrication of the chalcopyrite absorber layers
is based on rapid thermal processing (RTP) of stacked elemental
layers.~\cite{probst01} The Schottky devices have been fabricated by thermal
deposition of a 50nm thick Cr film on top of the uncleaned \CIGSSe surface,
followed by a 200nm thick Au layer for mechanical
protection.~\cite{deibel02ieee} Accelerated lifetime tests under standardized
damp-heat (DH) conditions at 85{\degree}C ambient temperature and 85\% relative
humidity were performed for various time spans ranging from 6h to
438h.~\cite{deibel01eupvsec} A subset of the Schottky devices was subject to
the DH test as well, depositing the metal front contacts only after DH
treatment of the absorber layer. The electrical characterization of the test
cells was done with the help of current--voltage characteristics, admittance
spectroscopy (using a Solartron 1260 impedance analyzer) and DLTS (using a
Semitrap 82E spectrometer). Temperature-dependent measurements in the range of
20K to 350K were carried out in a liquid Helium closed-cycle cryostat.

\begin{figure}
	\includegraphics[width=7.5cm]{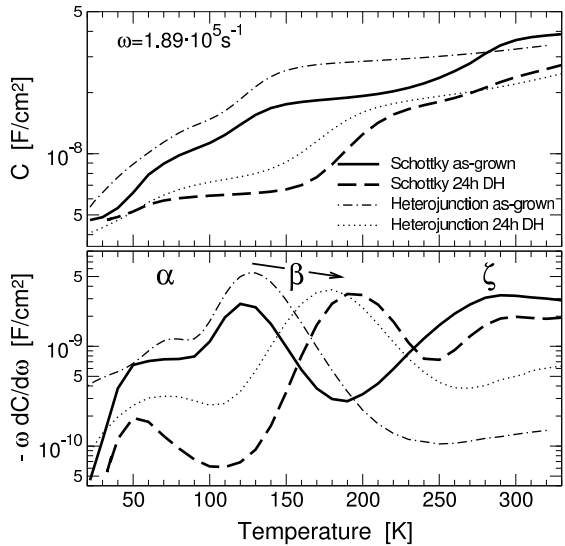}%
	\caption{Capacitance and differentiated capacitance (with respect to
			frequency) spectra versus temperature of the as-grown and 24h
			DH-treated Cr/\CIGSSe Schottky diodes. The corresponding admittance
			spectra of ZnO/CdS/\CIGSSe solar cells are juxtaposed for
			comparison. $\alpha$ can be attributed to the activation of the
			absorber layer, $\beta$ is characterized in the text. The deep trap
			$\zeta$ was discussed elsewhere.~\cite{deibel02ieee}
			\label{fig:ct+ct-diff-schottky}}
\end{figure}

\begin{figure}
	\includegraphics[width=7.5cm]{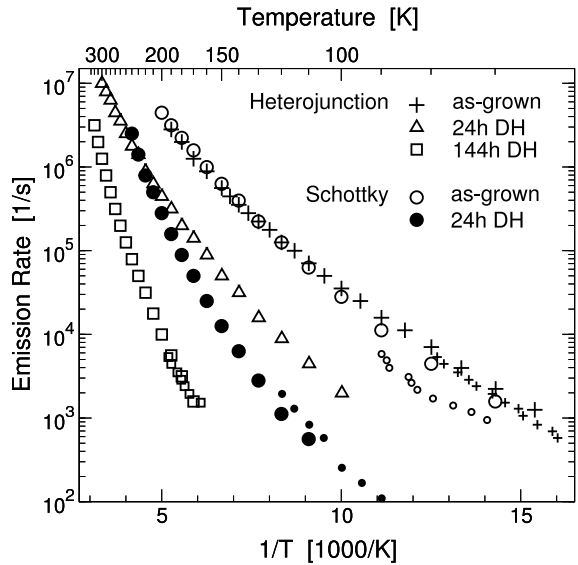}%
	\caption{Arrhenius plot of the emission rates of the defect state $\beta$,
			detected with admittance spectroscopy (larger symbols) as well as
			DLTS (smaller symbols) on both as-grown and 24h DH-treated
			Cr/\CIGSSe Schottky diodes and ZnO/CdS/\CIGSSe solar cells
			(as-grown and 24h/144h DH-treated), respectively. In both sample
			configurations, $\beta$ shows almost identical emission rates in
			the as-grown state and clearly shifts due to the DH treatment. The
			emission rates were extracted from differentiated capacitance
			spectra and from DLTS spectra recorded with a quiescent bias of 0V,
			applying filling pulses of 0.5V height and 20$\mu$s width.
			\label{fig:arrh-schottky-beta}}
\end{figure}

Upon applying capacitance spectroscopy, we disclose several different bulk
traps and a defect state, further on called $\beta$, in the ZnO/CdS/\CIGSSe
solar cells. The latter is commonly interpreted as an interface defect state
being located at the CdS/\CIGSSe heterojunction.~\cite{herberholz98,deibel01}
We were able to identify this defect state in our samples with admittance
spectroscopy as well as minority, majority, and reverse DLTS, respectively.
Preliminary results have been reported
elsewhere.~\cite{deibel01,deibel01eupvsec,deibel02ieee} The corresponding
admittance spectra (measured at 0V bias) of as-grown and DH-treated samples are
shown in Fig.~\ref{fig:ct+ct-diff-schottky}. The temperature-dependent emission
rates of the charge carriers from the defect state $\beta$ are summarized in an
Arrhenius representation (see Fig.~\ref{fig:arrh-schottky-beta}). In DLTS,
$\beta$ is extracted from a minority-carrier signal which can be measured
regardless of applying injection pulses or not. The activation energy of the
defect state $\beta$ increases continuously with time elapsed under DH
conditions from about 80meV in as-grown samples to about 340meV after 438h
exposure to DH conditions.~\cite{deibel01}  We have discovered the defect state
$\beta$ in the Cr/\CIGSSe Schottky junctions as well, for both, as-grown and
DH-treated absorbers. The corresponding admittance spectra are displayed in
Fig.~\ref{fig:ct+ct-diff-schottky}, the Arrhenius plots in
Fig.~\ref{fig:arrh-schottky-beta}. In the Schottky junctions, the typical shift
of the defect state $\beta$ with DH treatment to higher activation energies is
clearly recognized again. 

Note that the state $\beta$ can be commonly observed in ZnO/CdS/\CIGSSe solar
cells. Herberholz et al.~\cite{herberholz98} reported $\beta$ (referred to as
N1 therein) in ZnO/InS$_\textrm{x}$/\CIGSe heterostructures, we previously
detected it in ZnO/ZnSe/\CIGSSe heterojunctions and Cr/\CIGSSe Schottky
diodes.~\cite{deibel02ieee}  In the latter case, it is remarkable that the
emission rate of $\beta$ nearly coincides for as-grown heterostructure samples
and as-grown Schott\-ky diodes (see Fig.~\ref{fig:arrh-schottky-beta}). Our
experimental results provide evidence that the defect state $\beta$ is not
affected by the ZnO front contact (including the CdS or ZnSe buffer layer) and,
hence, located in the absorber layer.

\begin{figure}
	\includegraphics[width=6cm]{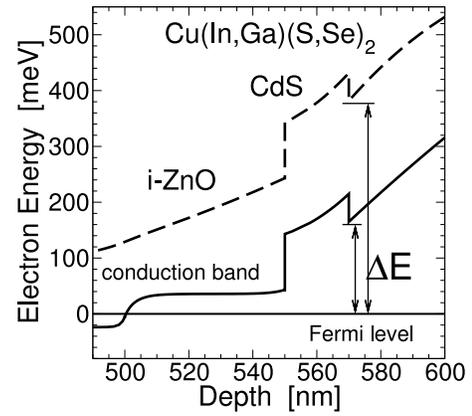}%
	\caption{A detail of the band diagram of a ZnO/CdS/\CIGSSe heterojunction
			solar cell. The assumed origin of the activation energy $\Delta E$
			of the interface defect state $\beta$ is sketched before (solid
			line) and after DH exposure (dashed line).\label{fig:band-detail}}
\end{figure}

The common interpretation of $\beta$ as energetical distribution of interface
defect states~\cite{herberholz98,deibel01} implies that the activation energy
equals the difference of conduction band minimum and electron quasi Fermi level
at the CdS/\CIGSSe interface. The assumed origin of this activation energy is
depicted in a blow-up of the corresponding band diagram in
Fig.~\ref{fig:band-detail}. Igalson et al.~\cite{igalson03} propose another
explanation for the origin of the defect state $\beta$. They hold four discrete
donor-like defect states in the absorber bulk (with activation energies of
80meV, 150meV, 350meV, and 570meV) responsible for the different instances of
$\beta$. They also report on an additional influence of the junction electric
field on the emission rates of those traps. 

We emphasize two experimental findings indicating an energy distribution of
states rather than the existence of discrete ones. First, the good agreement of
our different measurements with one distinct Meyer-Neldel rule points to a
common origin of the different instances of the defect state
$\beta$.~\cite{deibel01eupvsec} Second, the height of the capacitance step
changes continuously and is proportional to the activation
energy,~\cite{deibel01eupvsec} which would not be expected for the case of
several discrete defect states.

With the help of DLTS, we clearly identify $\beta$ as a minority-carrier
signal,~\cite{deibel02ieee} i.e., the width of the space charge region
diminishes directly after having applied the filling pulse. If $\beta$ were a
bulk trap, it would probably be a donor-like defect state. A prerequisite to
detect such a shallow bulk minority-carrier trap --- the activation energy of
$\beta$ for as-grown samples amounts to about 80meV --- is a large band bending
close to the interface between the front contact and the absorber, confining
the possible spatial location of the detection of the defect state to the
\CIGSSe surface. The crucial question remains why the above minority-carrier
defect state \emph{can} be detected using DLTS without the intentional
injection of minority carriers (either by optical injection or, in case of the
heterojunction, by injection pulses). For Schottky junctions based on n-Si, the
observation of minority-carrier traps with DLTS has been reported in case of
relatively large barrier heights.~\cite{stolt85} Comparable conditions are
given in \CIGSSe-based heterojunctions, because the observation of a type
inversion at the (in-vacuo) \CIGSe surface~\cite{schmid96} indicates a high
band bending. Thus, bulk minority-carrier defect states can be detected in
\CIGSSe heterojunctions and Schottky contacts even without minority-carrier
injection.

We consider the interpretation of $\beta$ as energy-distributed defect state in
the \CIGSSe surface region to be more likely than the proposition of Igalson et
al.~\cite{igalson03}. However, we can draw conclusions about the band bending
of the junction using either model. With the common interpretation of $\beta$
as an interface defect state, the information on the band bending is directly
obtained by the activation energy measured. In the Igalson model, the detection
of a minority-carrier defect state located in the absorber layer with
activation energies as low as 80meV is only possible for the electron quasi
Fermi level being very close to the conduction band. We conclude that the
detection of $\beta$ in as-grown \CIGSSe-based heterojunctions and Schottky
contacts signifies a strong band bending.

\CIGSSe films exposed to ambient air display a relatively small band bending,
as reported by Weinhardt et al.~\cite{weinhardt02} who measured a Fermi-level
position (relative to the conduction band minimum) of 0.5eV via applying
inverse photo\-emission spectroscopy. In-vacuo as-grown \CIGSe surfaces,
however, show a type inversion,~\cite{schmid96} i.e., the Fermi level is
located very close to the conduction band.~\cite{klein99} The impact of ambient
air on the band bending of \CIGSe films has been measured directly by Rau et
al.~\cite{rau99b} They observed a 200meV decrease of the band bending using
in-situ ultraviolet photoelectron spectroscopy after contact with ambient air
compared to the in-vacuo result. Devices finished after an air exposure of the
absorber layer, however, show a strong band bending
again.~\cite{deibel01eupvsec} In order to explain such a nontrivial phenomenon,
Rau et al.~\cite{rau99b} proposed that the CdS bath deposition is responsible
for a reintroduction of positive surface charges and a restauration of the band
bending to the state before the air exposure.

Our capacitance measurements demonstrate a strong band bending for both,
ZnO/CdS/\CIGSSe heterojunctions and Cr/\CIGSSe Schottky contacts, even though
the respective absorber layers were exposed to ambient air during processing.
Thus, the restauration of the band bending does not depend on the subsequent
CdS bath deposition, dissenting the previous model by Rau et al.~\cite{rau99b}
The concrete mechanism being responsible for such behavior is still unknown.
One possible explanation would be a Fermi-level pinning by a high concentration
of interface defect states developing only after contact formation, independent
of the contact layer applied, i.e., a reorganization of the chalcopyrite
surface due to the junction built-in electric field. Such an effect could
result from the field-induced migration of Cu.~\cite{rau99b} 

In summary, we have presented admittance spectroscopy and DLTS measurements on
ZnO/CdS/\CIGSSe heterojunctions and Cr/\CIGSSe Schottky contacts. A commonly
reported defect state with a continuously shifting activation energy depending
on the sample treatment presumably originates from an energy distribution of
defect states located at the \CIGSSe surface. The activation energy of this
energy distribution indicates the prevailing degree of band bending, which is
relatively small after air exposure, but increases considerably after the
formation of a Schottky contact or a heterojunction. The increase of the band
bending after the contact formation does not depend on the existence of the CdS
buffer layer.

\begin{acknowledgments}

The authors would like to thank J.~Palm and F.~Karg (Shell Solar, Munich)
for interesting discussions and for providing the heterojunction samples and
absorber layers. Fruitful discussions with B.~Dimmler, M.~Igalson, U.~Rau,
H.-W.~Schock and the partners of the Shell Solar joint research project at the
University of W{\"u}rzburg and the Hahn Meitner Institute Berlin are also
acknowledged.

\end{acknowledgments}


\bibliography{journale,Deibel}

\end{document}